# ActiveAI: Introducing AI Literacy for Middle School Learners with Goal-based Scenario Learning

Ying Jui Tseng, Gautam Yadav
Carnegie Mellon University
yingjuit@andrew.cmu.edu

## Brief Abstract

The ActiveAI project addresses key challenges in AI education for grades 7-9 students by providing an engaging AI literacy learning experience based on the AI4K12 knowledge framework. Utilizing learning science mechanisms such as goal-based scenarios, immediate feedback, project-based learning, and intelligent agents, the app incorporates a variety of learner inputs like sliders, steppers, and collectors to enhance understanding. In these courses, students work on real-world scenarios like analyzing sentiment in social media comments. This helps them learn to effectively engage with AI systems and develop their ability to evaluate AI-generated output. The Learning Engineering Process (LEP) guided the project's creation and data instrumentation, focusing on design and impact. The project is currently in the implementation stage, leveraging the intelligent tutor design principles for app development. The extended abstract presents the foundational design and development, with further evaluation and research to be conducted in the future.

## The Learning Engineering Process Model

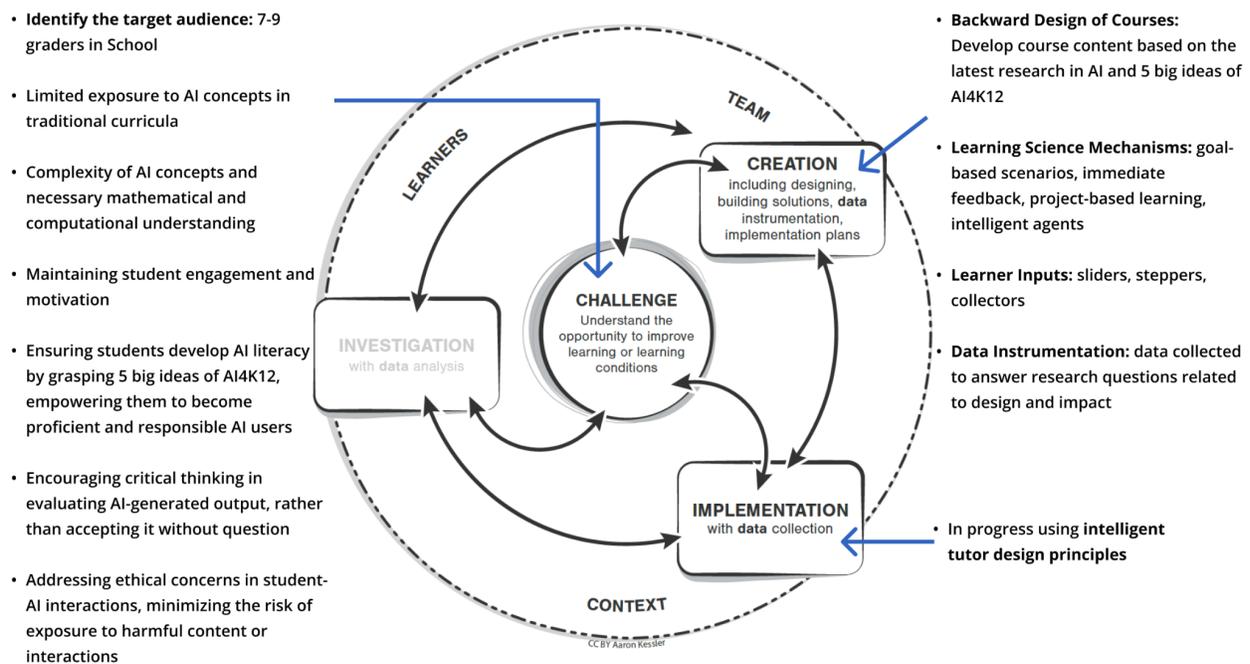

Figure 1. Visualizing the Learning Engineering Process: A Comprehensive Model for Designing and Implementing ActiveAI (Kessler et al., 2022)

# Extended Abstract

ActiveAI project offers 7-9 grade students an engaging and accessible AI learning experience, grounded in AI4K12's five big ideas (Touretzky et al., 2022). These five big ideas include "Perception," "Representation & Reasoning," "Learning," "Natural Interaction," and "Societal Impact." The objective is to foster an understanding of AI concepts and applications in an ever-changing technological landscape.

**Challenge Stage:** The challenges the project seeks to address include:
1. Limited exposure to AI concepts in traditional curricula.
2. Complexity of AI concepts and necessary mathematical and computational understanding.
3. Maintaining student engagement and motivation.
4. Ensuring students develop AI literacy by grasping 5 big ideas of AI4K12, empowering them to become proficient and responsible AI users.
5. Encouraging critical thinking in evaluating AI-generated output, rather than accepting it without question.
6. Addressing ethical concerns in student-AI interactions, minimizing the risk of exposure to harmful content or interactions.

**Creation Stage:** ActiveAI Creation Step is designed to teach 7-9 grade school students the five big ideas of AI4K12 effectively. The design of ActiveAI relies on proven learning science mechanisms and a range of learner inputs to engage students and promote in-depth understanding.

1. Goal-based scenarios: ActiveAI adopts a goal-oriented approach, offering a range of scenarios in which students tackle real-world challenges that call for AI principles and methodologies (Schank et al., 1994). By immersing learners in simulated situations that present them with a clear objective, they are incentivized to invest more effort into grasping fundamental concepts and honing their skills.
   As an example, students take on the role of content creators seeking to gauge the sentiment of comments on their TikTok videos. They leverage AI techniques to develop a sentiment analysis model, which they utilize to gain insights into their audience's reactions and produce more engaging content.
   In another module, students engage in a task of classifying dog images, much like an automated image tagging system that makes search results more precise in platforms like Instagram. They utilize supervised image classification techniques to label images as "dog sleeping" or "dog playing". During this process, they face the challenge of a skewed training dataset where most images show dogs sleeping indoors and playing outdoors. This could lead to the learning of irrelevant features and biases in the results. Through this interactive task, students not only delve into the workings of machine learning models, but they also learn about potential biases that can emerge from unbalanced datasets, and the importance of addressing these issues to ensure accurate and fair outcomes.
2. Immediate feedback: Throughout the learning experience, students receive immediate feedback on their performance through the app's intelligent agents and the learner inputs.
3. Project-based learning: ActiveAI encourages students to work on authentic hands-on projects that require the application of AI concepts and skills.

4. Intelligent Agents: The app utilizes intelligent agents that enable students to engage with real AI algorithms without the need for coding skills. This approach effectively reduces the learning barrier for students and enables them to access more advanced concepts that were previously beyond their reach (Ng et al., 2022). Studies have demonstrated that incorporating intelligent agents in educational settings can effectively engage middle school learners and promote the development of AI literacy skills (Rodríguez-García et al., 2020).
5. Learner input: ActiveAI strategically employs a select set of learner input interactions – sliders, steppers, and collectors – to facilitate student engagement and comprehension of complex AI concepts. By limiting the interactions to these three types, the app ensures a consistent and familiar interaction across the system, reducing the learning curve and allowing students to focus on mastering AI Literacy concepts without the need to constantly adapt to new interaction styles. The following descriptions detail each interaction type:
    a. Collector: Collector lets learners capture their own datasets or label provided dataset into different classes to train the intelligent agent. For example, students can use their tablet's camera to take pictures for a basic classification model. By participating actively in data collection, learners gain a deeper understanding of the process and a stronger connection to the problem. This hands-on experience enhances student engagement and reinforces the importance of data collection for real-world problem-solving.
    b. Slider: With sliders, learners can adjust variables and receive immediate feedback on the screen. Sliders can be employed to modify various quantitative variables, such as the threshold for a classification model or the amount of training data used. This real-time feedback fosters deeper understanding and experimentation as students explore AI literacy concepts.
    c. Stepper: Steppers are designed to control time-related variables or provide learners with step-by-step explanations. By offering a guided approach to navigating complex AI processes, steppers help students grasp concepts sequentially and methodically.

Figure 2 describes how the collector and slider works in the dog classification scenario.
Figure 3 describes how the stepper works in the TikTok sentiment analysis scenario.

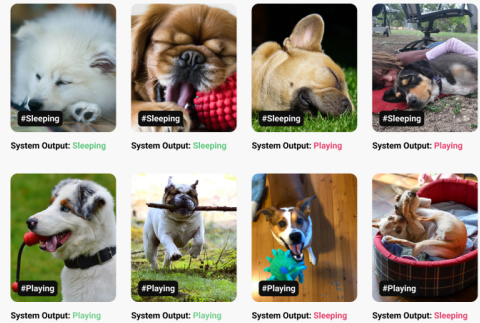

Figure 2. Students employ a collector to categorize 10 images into two classes - "sleeping" and "playing" to create a training set for a classifier (left side). In another learning activity, they will manipulate a slider to adjust the number of training examples, thereby observing how this modification influences the accuracy of their automated tag system on a given test dataset of 8 images (right side).

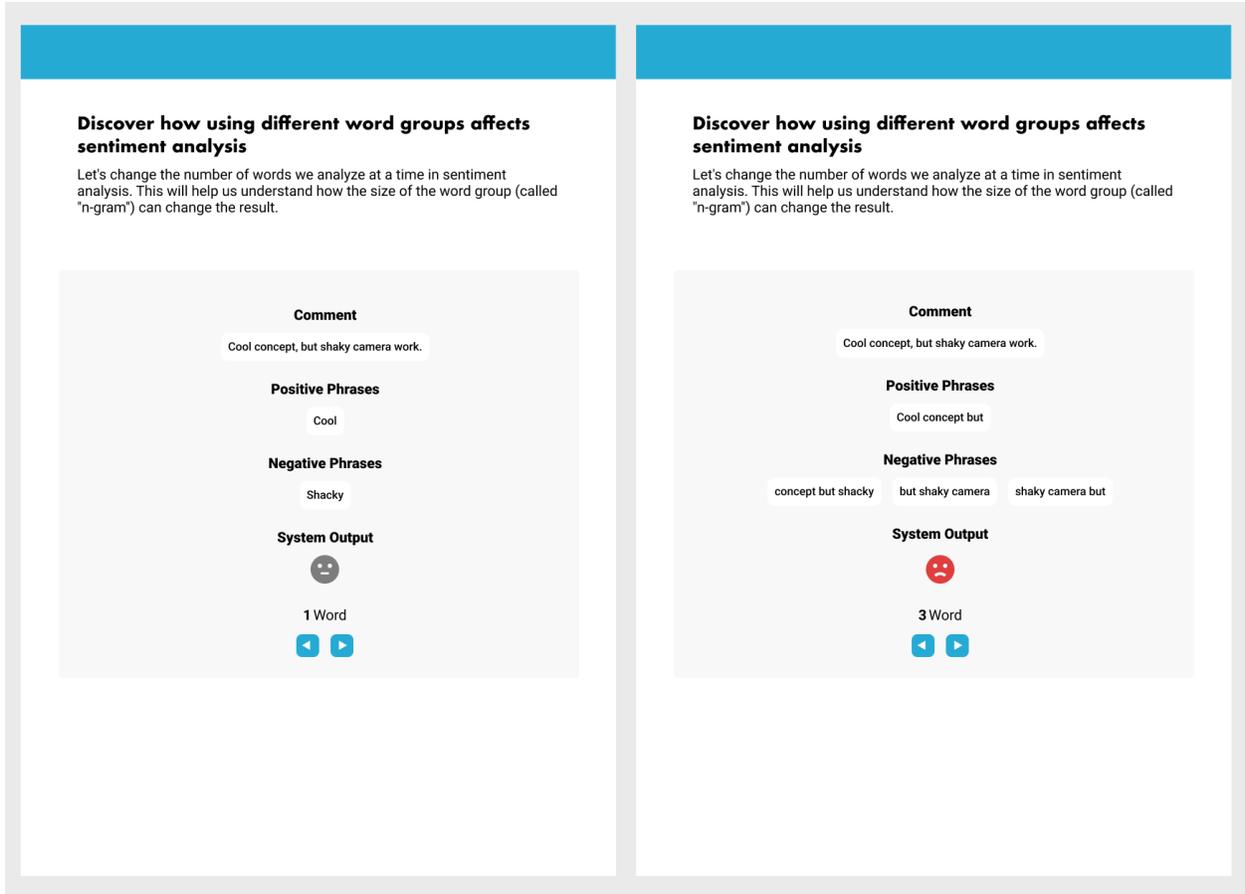

Figure 3. Students use a stepper to modify the number of words considered in the sentiment analysis, enabling them to explore how using different word groups affects sentiment analysis results sequentially and gain insight into how machines "understand" sentiment.

The data collected from students' app interactions will be analyzed to answer research questions related to design and impact (Torrence, 2022), providing insights into the learning experience's effectiveness and continuous improvement.

Table 1. Design Focus Data Instrumentation

| Research Questions | Hypothesis | Method | Metrics |
| --- | --- | --- | --- |
| How effectively are students engaging with the app and its different interactions? | Increased use of interaction types leads to better understanding and problem-solving. | Analyzing frequency and type of interactions | Number of interactions, interaction types, time spent on each interaction |
| How do the hints provided by the tutor affect student performance? | Hints offered by tutor will positively impact student performance by | Analyzing hint usage and impact on performance | Number of hints used, performance before and after hint usage |

| | providing appropriate guidance. | | |
|---|---|---|---|
| How does the level of assistance provided by the tutor correlate with student performance? | A tailored level of assistance will result in improved student performance compared to generic hints. | Correlating assistance level with student performance | Assistance level, student performance |

Table 2. Impact Focus Data Instrumentation

| Research Questions | Hypothesis | Method | Metrics |
|---|---|---|---|
| How quickly are students progressing through the app's modules and activities? | Students who spend more time on the app will demonstrate faster progress through modules and activities. | Tracking student progress through modules and activities | Time spent on app, number of modules completed, time taken to complete activities |
| What factors contribute to students' time taken to complete each activity or module? | Factors such as prior knowledge, level of tutor assistance, and engagement will impact completion times. | Analyzing time taken and factors influencing completion times | Time taken to complete activities, prior knowledge, engagement, tutor assistance |
| How does student performance on assessments change after using the app? | Student performance on assessments will improve after using the AI literacy app. | Comparing pre- and post-app assessment scores | Pre- and post-app assessment scores |
| Which components of the app have the most significant impact on assessment scores? | Hands-on activities and personalized feedback will have the most significant impact on improving scores. | Analyzing correlation between app components and assessment scores | App component usage, Assessment scores |
| How does student performance on assessments related to the same KC change over multiple opportunities? | As students have more opportunities to practice the same KC, their performance on related assessments will improve. | Analyzing assessment performance over multiple opportunities | Number of opportunities, assessment scores for the same KC, student progress |
| How does the performance of students using the intelligent agent compare to those using pre-defined practice questions? | Students using the intelligent agent will show better performance compared to those using pre-defined practice. | Comparing performance between intelligent agent users and pre-defined practice question users | Assessment scores, usage of intelligent agent, usage of pre-defined practice questions |

**Implementation Stage:** The ActiveAI project is currently in the implementation stage, utilizing the intelligent tutor design principles (Corbett et al., 1997) to develop an interactive learning environment that incorporates the learning science mechanisms and learner inputs discussed in the Creation Step (Aleven et al., 2006). By leveraging intelligent tutor design principles, we can build an adaptive and personalized learning experience, ensuring students receive appropriate support and guidance throughout their AI literacy journey.

This extended abstract presents the foundational design and development of the ActiveAI program, with further evaluation and research to be conducted in the future.